\documentclass[]{interact}
\usepackage{amssymb}
\usepackage{amsmath}
\usepackage{hyperref}
\hypersetup{
    colorlinks=true,
    linkcolor=cyan,
    filecolor=cyan,
    urlcolor=cyan,
    citecolor=cyan,
}
\usepackage{natbib}
\bibpunct[, ]{(}{)}{,}{a}{}{,}
\usepackage{booktabs}
\usepackage{siunitx}
\usepackage{setspace}
\usepackage{booktabs}
\usepackage{makecell}
\usepackage{tabularx}
\usepackage{graphicx}
\usepackage{longtable}
\usepackage{threeparttable}
\usepackage{ltablex}
\usepackage{ragged2e}
\usepackage{adjustbox}
\usepackage{float}
\usepackage{tocloft}
\usepackage[utf8]{inputenc}
\usepackage[justification=centering]{caption}
\usepackage{xcolor,pifont}
\usepackage{graphicx}
\usepackage{subcaption}
\newcommand{\circfour}[1]{
  \begingroup
  \def\full{\textcolor{black}{\ding{108}}}
  \def\empty{\textcolor{black!25}{\ding{108}}}
  \ifcase#1 \empty\empty\empty\empty
  \or      \full \empty\empty\empty
  \or      \full \full \empty\empty
  \or      \full \full \full \empty
  \else    \full \full \full \full
  \fi
  \endgroup
}
\usepackage{setspace}
\usepackage{amssymb}
\usepackage{amsmath}
\usepackage{hyperref}
\hypersetup{
    colorlinks=true,
    linkcolor=cyan,
    filecolor=cyan,
    urlcolor=cyan,
    citecolor=cyan,
}

\usepackage{setspace}

\begin{document}
    \title{Unbox Responsible GeoAI: Navigating Climate Extreme and Disaster Mapping}
    
    \author{
        \name{Hao Li\textsuperscript{a},\thanks{CONTACT Hao Li. Email: hao.li@nus.edu.sg}
        Steffen Knoblauch\textsuperscript{b,c,d}}
        \affil{
        \textsuperscript{a}Department of Geography, National University of Singapore, Singapore; 
        \textsuperscript{b} GIScience Research Group, Heidelberg University, Germany; 
        \textsuperscript{c} HeiGIT at Heidelberg University, Heidelberg, Germany;
        \textsuperscript{d} Interdisciplinary Centre of Scientific Computing (IWR), Heidelberg University, Heidelberg, Germany;
        }
    }
    \maketitle
    \begin{abstract}

    As climate extreme and disaster events become more frequent and intense, Geospatial Artificial Intelligence (GeoAI) has emerged as a transformative approach for large-scale disaster mapping and risk reduction. However, the purely mechanical, performance-driven deployment of GeoAI models can result in amplifying inherent spatial inequalities, preventing effective emergency decision-making, and producing severe environmental carbon footprint. To unbox the concept of responsible GeoAI, this position paper  examines its emerging role, e.g., in climate extreme and disaster mapping, from a critical GIS perspective. We address the nexus of responsible GeoAI into four interrelated theoretical dimensions, specifically Representativeness, Explainability, Sustainability, and Ethics, with examples from climate extreme and disaster mapping. Moreover, targeting at the operational practice, we then propose a conceptual governance Model of responsible GeoAI that categorizes its governance practices into Data, Application, and Society scopes. Last, this position paper aims to raise the attention in the broader GIS community that the future of climate resilience relies not just on building better algorithms, but on fostering a governance ecosystem where GeoAI is deployed responsibly, ethically, and sustainably.
 
        \end{abstract}




\section{Introduction} \label{intro} 

As Earth’s climate changes, it is impacting climate extreme events and disasters across the planet. Record-breaking heatwaves, drenching rainfalls, extreme wildfires, and widespread flooding during hurricanes are all becoming more frequent and more intense. Compared with the average for the last 30 years (i.e., 1994-2023), the major natural disasters in 2025 showed an 8\% increase in frequency, an 74\% increase in fatalities, an 42\% more affected population, and an 49\% increase in direct economic losses based on the EM-DAT statistic \citep{teber2025geo}. This trend post increased pressure on global communities, where limited resources struggle to keep pace with growing damages, contributing to the Disaster Resilience Gap, which is further exacerbated by the digital divide between the Global North and South countries \citep{ghosh2023very,hallegatte2019lifelines}. 

Geospatial Artificial Intelligence (GeoAI) is rapidly transforming the ways how geographers map, study, and engage with the social place and physical environment across the world \citep{openshaw1997artificial, li2024geoai}. The advances in GeoAI has already expanded the methodological toolkit of geography, offering capabilities for multimodal data integration, large-scale mapping, and predictive spatiotemporal modeling. More recently, Generative AI (GenAI) and Foundation Models (FMs), in particular, open up possibilities for across data modalities in geographical data analysis \citep{wang2024gpt, mai2024opportunities, janowicz2025geofm}, including remote sensing Earth Observation (EO) data, ground-level Street View Imagery (SVI), and crowdsourced Volunteered Geographical Information (VGI) \citep{li2026synergizing}. In this context, GeoAI has become a transformative force in climate extreme and disaster mapping \citep{zou2023geoai, li2025cross}. However the current advances most focus on the efficiency and effectiveness of the technical approaches, leaving the society and humanistic consideration substantially lagged.

Despite the promise of the integration of GeoAI approaches into climate extreme and disaster mapping at scale, if developed or deployed irresponsibly, they may result in widening spatial inequalities, misrepresenting marginalized communities, and generating unsustainable environmental costs, etc \citep{liu2025sustainable, shi2025geography}. While GeoAI approaches frequently achieve unprecedented predictive accuracy, e.g., in detecting hazard footprints or classifying damaged infrastructure, their explainability and interoperability remains fundamentally questioned. In critical application domains, such as disaster management, predictive accuracy alone is an insufficient metric for responsible operational decision-making. Thus, the black-box nature of these models violates the core professional responsibility of geographer and stakeholders to provide justifiable, responsible, and transparent evidence for disaster response policy making that directly dictate resource allocation and human survival. Therefore, Responsible GeoAI plays a vital role when it comes to naturally-sophisticated geographical questions (e.g., climate extreme and disaster mapping), where the ethics, fairness, inclusivity, and responsibility of AI systems matters. 

\begin{figure}[!ht]
    \centering
    \includegraphics[width=0.8\linewidth]{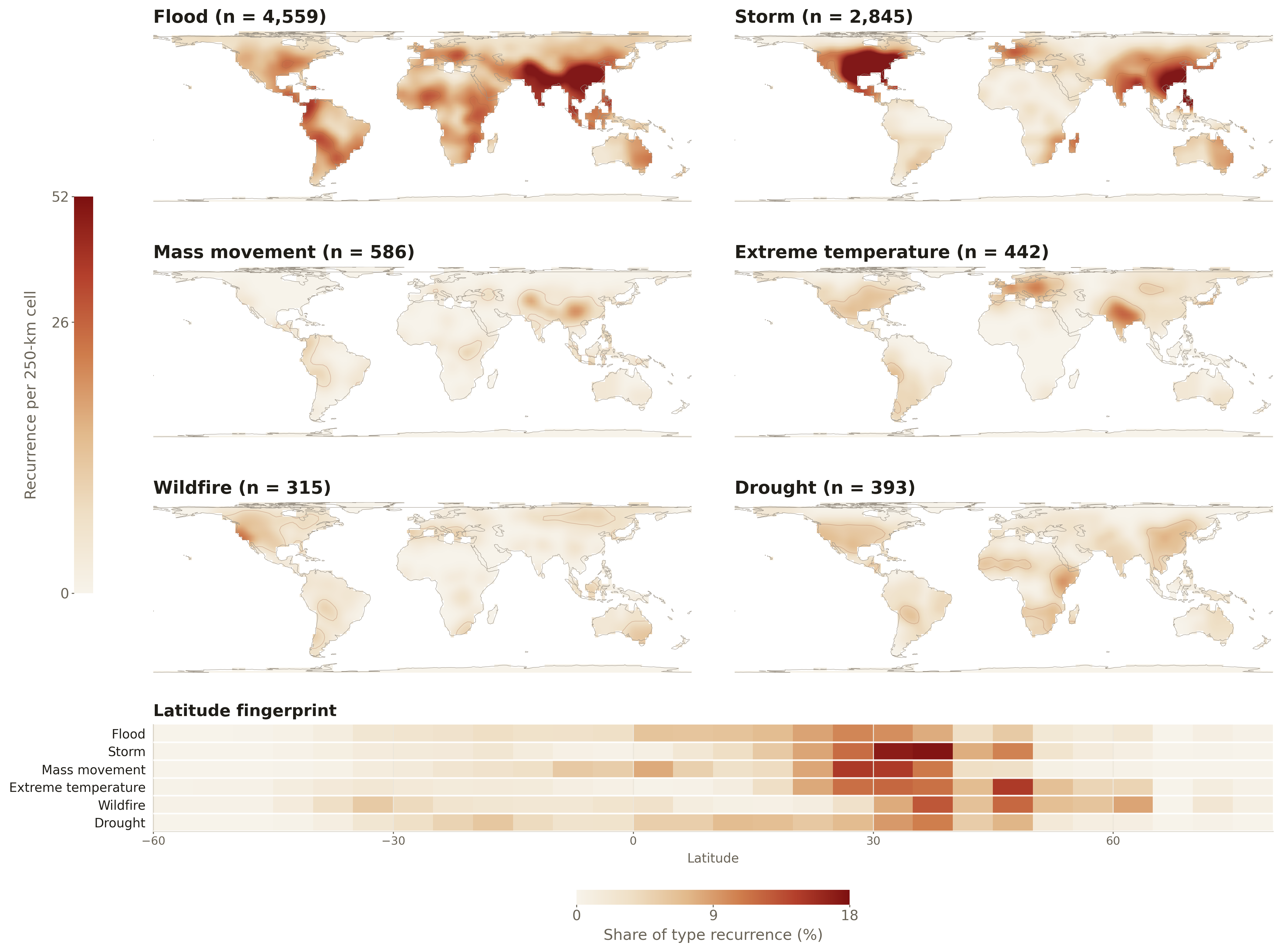}
    \caption{Geographical distribution of geocoded EM-DAT events by disaster type (1990–2023) \citep{teber2025geo}. Spatial pattern varies: floods are everywhere and most frequent; storms and mass movements cluster along storm belts and mountainous regions; and temperature-driven events exhibit strong latitudinal biases, with extreme temperatures spanning mid-to-high latitudes and droughts mainly in low latitudes.}
    \label{fig:diasater_map}
\end{figure}

\begin{figure}[!ht]
    \centering
    \includegraphics[width=0.8\linewidth]{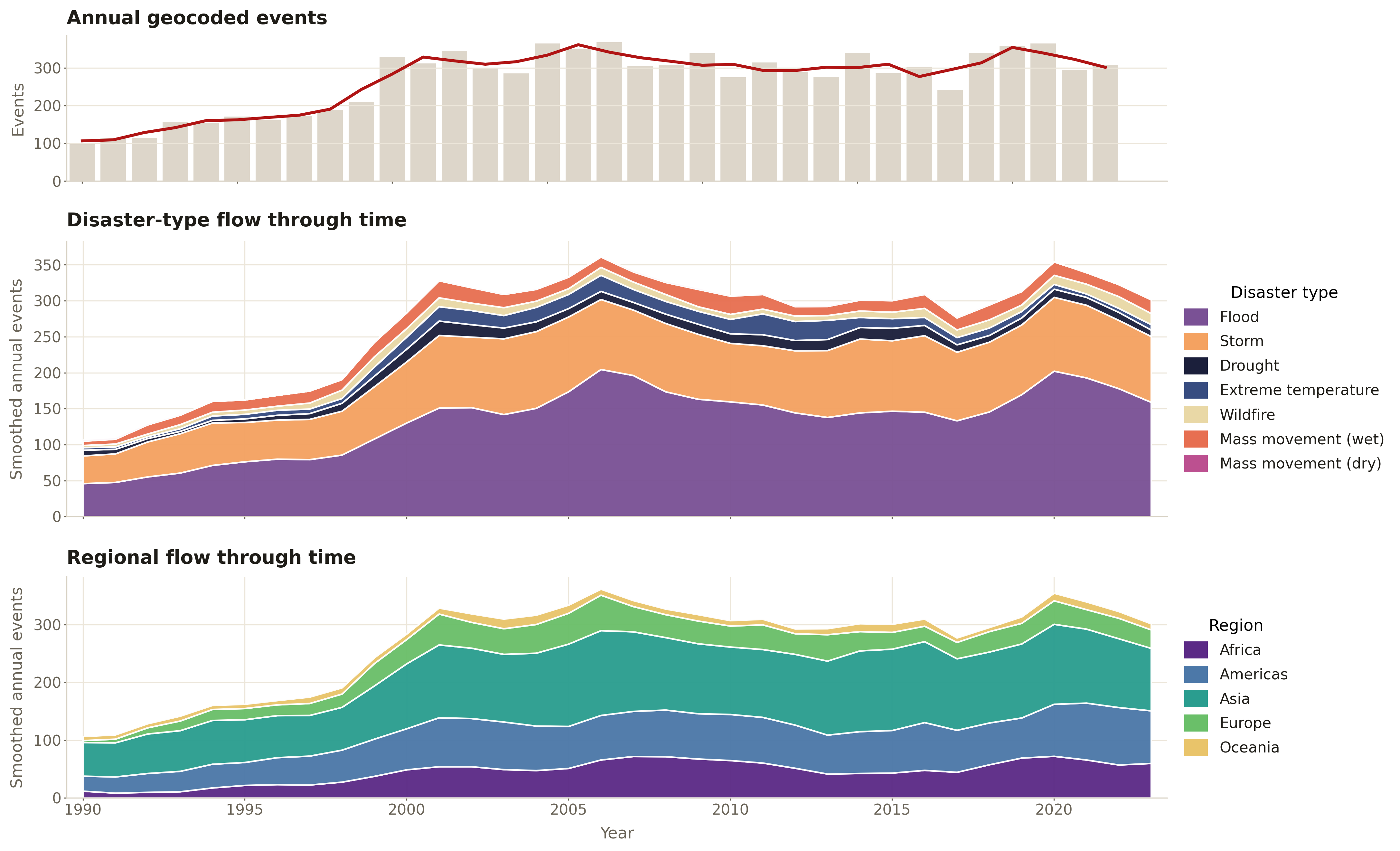}
    \caption{Annual distribution of geocoded EM-DAT events (1990–2023) reflects a general trend, with reported amount steadily increasing before stabilizing around the year 2000 \citep{teber2025geo}. Despite this pre-2000 variance, the 1990s still provide 100 to 200 high-quality event records per year, ensuring the dataset remains viable for robust longitudinal studies over the past three decades.}
    \label{fig:disaster_temporal}
\end{figure}

In this position paper, we aims to address the emerging role of responsible GeoAI for climate extreme and disaster mapping, by critically thinking  its transformative potentials and underlying risks from a GIS perspective. First, we analysis the nexus of responsible GeoAI by conceptualizing into four interrelated dimensions: 1) \textbf{Representativeness}, 2) \textbf{Explainability}, 3) \textbf{Sustainability}, and 4) \textbf{Ethics}. Building upon these dimensions, we then design a conceptual governance Model of Responsible GeoAI from the lens of climate extreme and disaster mapping. Finally, we share our perspectives about the key challenges and opportunities for the integration of this governance Model into real-world disaster risk reduction workflows, for instance in Southeast Asia. Last but not the least, we argue that shaping a disaster-resilient future demands that the GIS community actively embed responsibility into the design and deployment of GeoAI systems, ensuring these technologies foster equity, sustainability, and trustworthy advances in a rapidly changing climate

\section{The Nexus of Responsible GeoAI}

By embedding principles of responsibility into the design, development, and governance of AI, we geographers strive to ensure that the advances in AI not only empower the analytical capacity of the discipline but also ensure responsible implications from both social and environmental perspectives. 

\begin{figure}[!ht]
    \centering
    \includegraphics[width=0.65\linewidth]{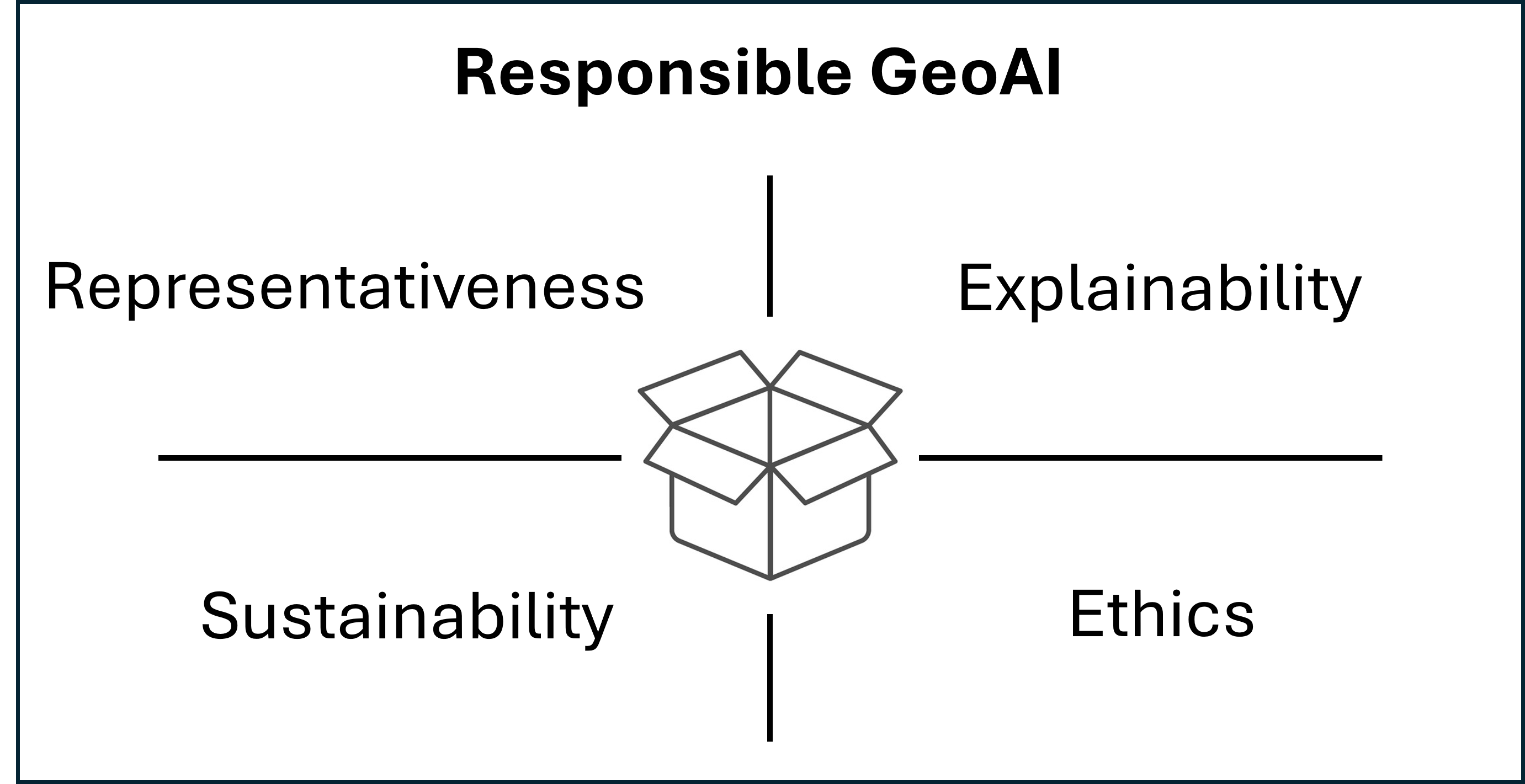}
    \caption{The Nexus of Responsible GeoAI featuring four interrelated dimensions: Representativeness, Explainability, Sustainability, and Ethics.}
    \label{fig:nexus}
\end{figure}

To facilitate this goal, as shown in Figure \ref{fig:nexus}, this position paper approach this ongoing debate by unboxing the nexus of responsible GeoAI by conceptualizing into four interrelated dimensions . From a GIS perspective, this paper offers a critical and forward-thinking reflection on how GeoAI shall be designed and deployed in climate extreme and disaster mapping tasks. 

\subsection{Representativeness}

For geographic data, absence is as profound as presence as introduced in \cite{sutton2026missing}. Herein, representativeness requires critically examining not only the spatial and demographic coverage of the data used to train and develop GeoAI models, but also the epistemological biases embedded within their ontologies \citep{herfort2023spatio, manvi2024large}. Actually, the term "Representativeness" was first examined in \citet{zhang2018representativeness}, where authors define the representativeness of geographic samples refer to the degree to which the spatial variation and attribute diversity are covered in the samples used to model the whole study areas, so called geographic sampling methods \citep{curtis2000approaches, yang2013integrative}. However, in the AI era, the representativeness is no longer limited to the samples but far early propagated into the model built and the predictions made, thus deserves a new reflections for its contemporary implications.

As one example and form of open-source geographic data, OpenStreetMap (OSM) sufferes from a severe Global North bias \citep{herfort2023spatio, huang2025geospatial}. This representativeness gap leaves vulnerable communities, low-income population, and indigenous territories digitally invisible and entirely absent. Moreover, the rationale for urgently addressing this is rooted in the high failure rate of GeoAI transferability and generalizability, e.g., when applying disaster mapping models across diver urban environments. For example, the GeoAI model trained primarily on the geometry and spectral features of building footprints in US or European cities will foreseeable fail to detect informal settlement in the densely packed, heterogeneously constructed areas of Sub-Saharan Africa \citep{li2022improving}.

\begin{figure}[!ht]
    \centering
    \includegraphics[width=0.8\linewidth]{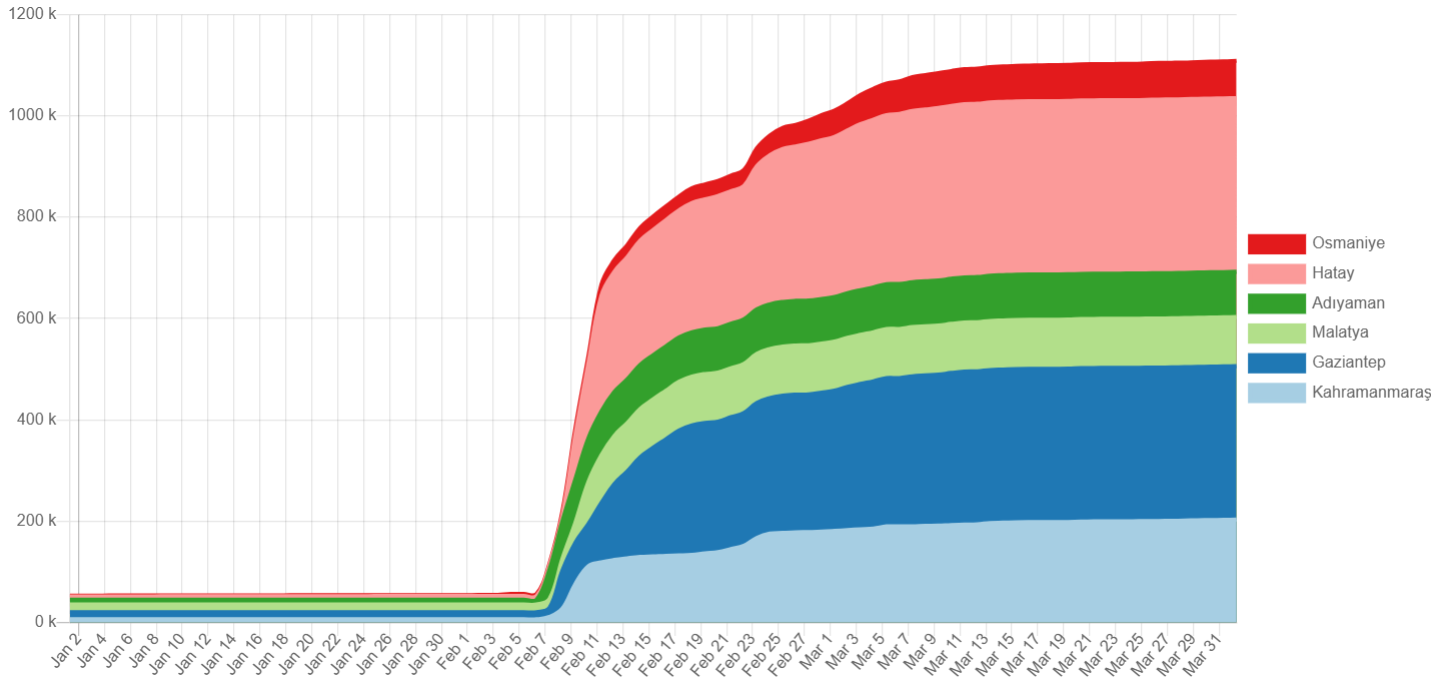}
    \caption{Daily temporal change patterns of OpenStreetMap building data over six damaged regions during the Earthquake in Syria and Türkiye in early 2023.}
    \label{fig:turkey_earthquake}
\end{figure}

For the mapping community, climate extreme or disaster events typically triggers an unprecedented explosion of generated data, ranging from rapid response satellite imagery to crowdsourced reports and editing. For instance, Figure \ref{fig:turkey_earthquake} shows the temporal changes in OSM buildings during the 2023 Turkey-Syria Earthquake. Visualization is created used the ohsome dashboard developed by HeiGIT gGmbH at Heidelberg University \footnote{https://dashboard.ohsome.org/}. However, the sheer volume of disaster mapping contributions does not directly contribute to better representativeness. In contrast, the opportunistic harvesting of big data during crises often introduces and leads to severe socio-demographic biases during the disaster mapping \citep{zhou2022victimfinder}. The core challenge for responsible GeoAI deployment in disaster decision-making lies in how to critically evaluate and maintain the representativeness of these massive datasets. For instance, relying on the automated extraction of location data from massive social media streams to map flood rescue requests inherently biases response efforts toward digitally connected, younger, or wealthier demographics \citep{fan2020spatial, zhou2022victimfinder}. This systematic skew renders elderly, impoverished, or socially marginalized populations remains digitally invisible during the disaster \citep{kawachi2020disaster}.

In this context, representativeness is considered a key dimension of responsible GeoAI and its deployment in disaster mapping scenarios. To effectively navigate through this dimension, modern GeoAI-based systems must be designed to empower, rather than replace, disaster mapping communities. Initiatives like the Humanitarian OpenStreetMap Team's (HOT) AI-assisted tools (e.g., fAIr) exemplify this collaborative paradigm, echoing existing theoretical Models such as the Centaur VGI \citep{huck2021centaur}. By integrating human-AI-in-the-loop concept, the safeguard of representativeness can guarantee that GeoAI models accurately and equitably serve the diverse, highly vulnerable geographies they are intended to protect.

\subsection{Explainability}
    
The inherent spatial dependency, spatial heterogeneity, and scale effects of geographical phenomena often challenge the black-box nature of AI systems \citep{kwan2012uncertain, jiang2015geospatial,zhu2018spatial,song2020optimal}. Recently, explainability have emerged as a central research themes in Ai, reflecting growing demands for reliable, transparent, and responsible decision-making system \citep{hu2024five, li2025explainable}. In climate extreme and disaster mapping, GeoAI models frequently operate on heterogeneous, multi-source spatial data, introducing uncertainty from data noise, temporal dynamics, model interpretability, and etc \citep{xing2023challenges, liu2024explainable, song2026geoai}. At the same time, the complex nature of deep models can lead to limited explainability. Therefore, to promote explainability is essential to promote responsible usage of GeoAI in critical applications such as disaster response and climate extreme mitigation.

\begin{figure}[!ht]
    \centering
    \includegraphics[width=0.9\linewidth]{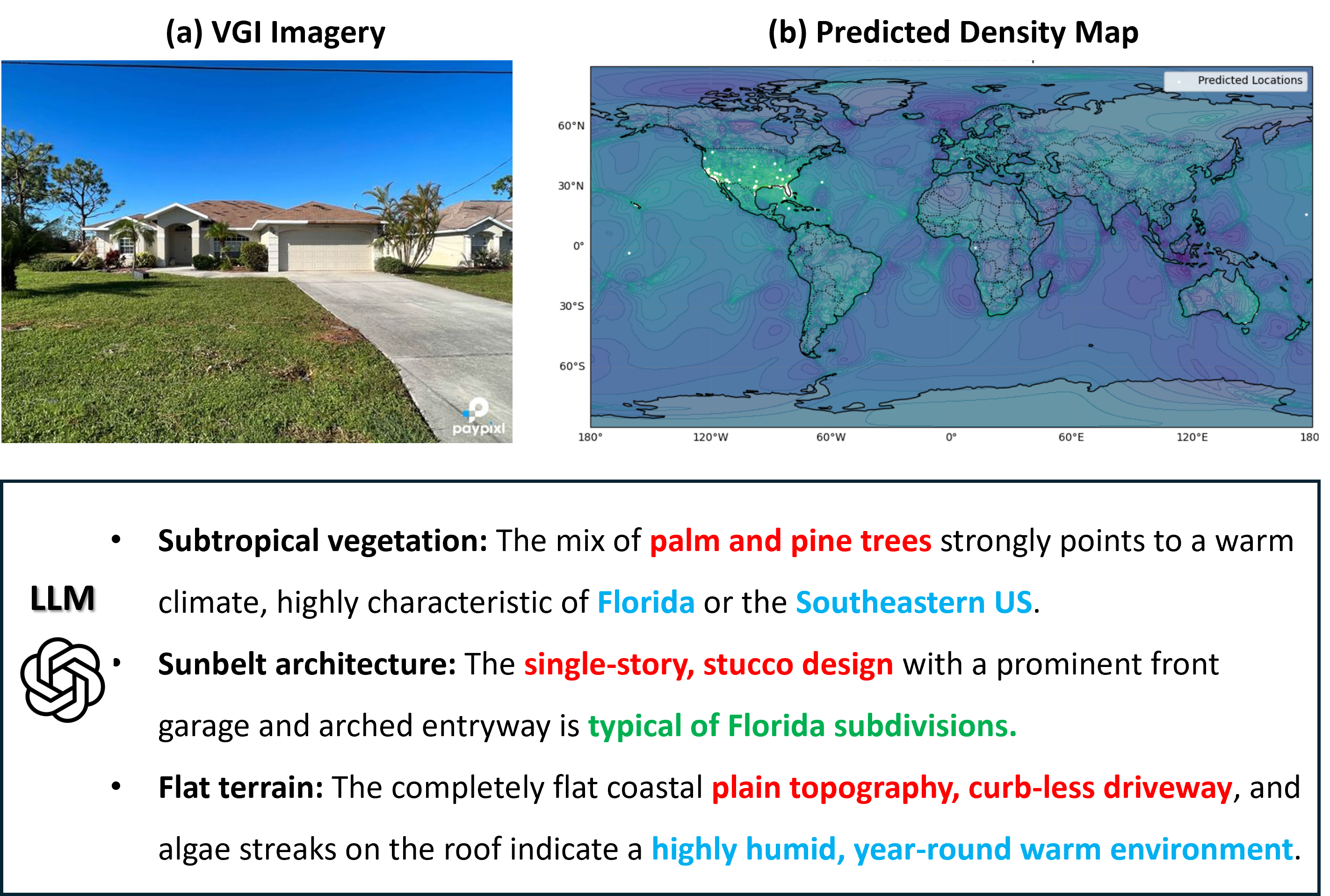}
    \caption{Examples of VGI imagery geolocalization using two GeoAI approaches, highlighting the different explainability strategies. (a) Crowdsourced SVI image from Florida, USA during the Hurricane IAN 2022; (b) the probabilistic density map generated by the ProbGLC method proposed in \citep{li2025towards} showing the potential predicted locations; (c) the geospatial reasoning provided by LLM (i.e., Gemini-3.5) in terms of geolocalization clues. }
    \label{fig:explainability}
\end{figure}

Explainable GeoAI requires that we unpack how fundamental spatial relationships, such as proximity, topology, spatial autocorrelation, and scale, are interpreted and resonated within predictive models. Recent advancements, for instance GeoShapley \cite{li2024geoshapley}, demonstrate how explainability can be achieved in common geographical analysis. More recently, advanced multimodality GeoAI approach, for instance vision-language models, shows great potential in critical tasks like probabilistic geolocalization \citep{li2025towards} and LLM-based geospatial reasoning \citep{ji2025foundation}, providing new avenues toward more Explainability in GeoAI model. Figure \ref{fig:explainability} give an example for how the vision- and language-based explainability approaches can complement each other in an VGI imagery geolocalization task.

In this regard, there is growing emphasis on developing intrinsically explainable models that integrate spatiotemporal knowledge, physical principles, and domain expertise, in the GeoAI models \citep{mai2024srl,li2024geoai}. The urgent need for domain-oriented explainability techniques is key to facilitate decision-making in time-critical disaster response scenarios.

\subsection{Sustainability}

The integration of GeoAI presents a profound ecological paradox. While it offers unprecedented tools for climate modeling and disaster mapping, the computational cost required to train massive Geospatial Foundation Models (FMs) generate a staggering energy footprint \citep{shi2025geography}. As the scale and complexity of GeoAI models continue to grow, sustainability has become an important dimension, especially as those models are increasingly adopted in large-scale and operational monitoring the global climate extreme and disasters \cite{van2021sustainable, jay2024prioritize}. In the context of responsible GeoAI, sustainability refers to the development and deployment of AI-based systems to minimize their environmental impact while promoting computational efficiency and social justice.

\begin{figure}[!ht]
    \centering
    \begin{subfigure}{0.48\textwidth}
        \includegraphics[width=\linewidth]{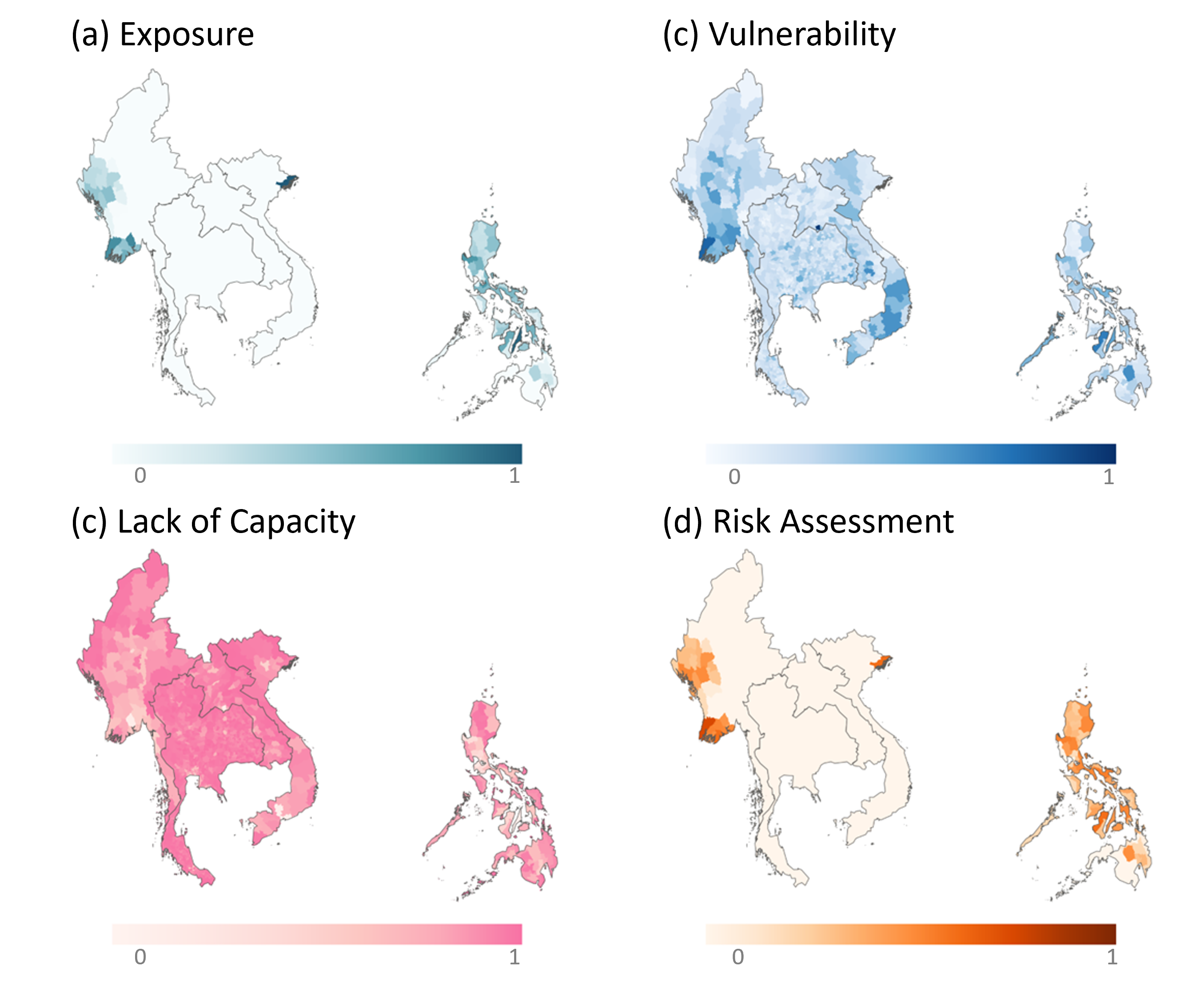}
        \caption{Cyclone risks in Southeast Asia Countries.}
        \label{fig:cyclone}
    \end{subfigure}
    \hfill
    \begin{subfigure}{0.48\textwidth}
        \includegraphics[width=\linewidth]{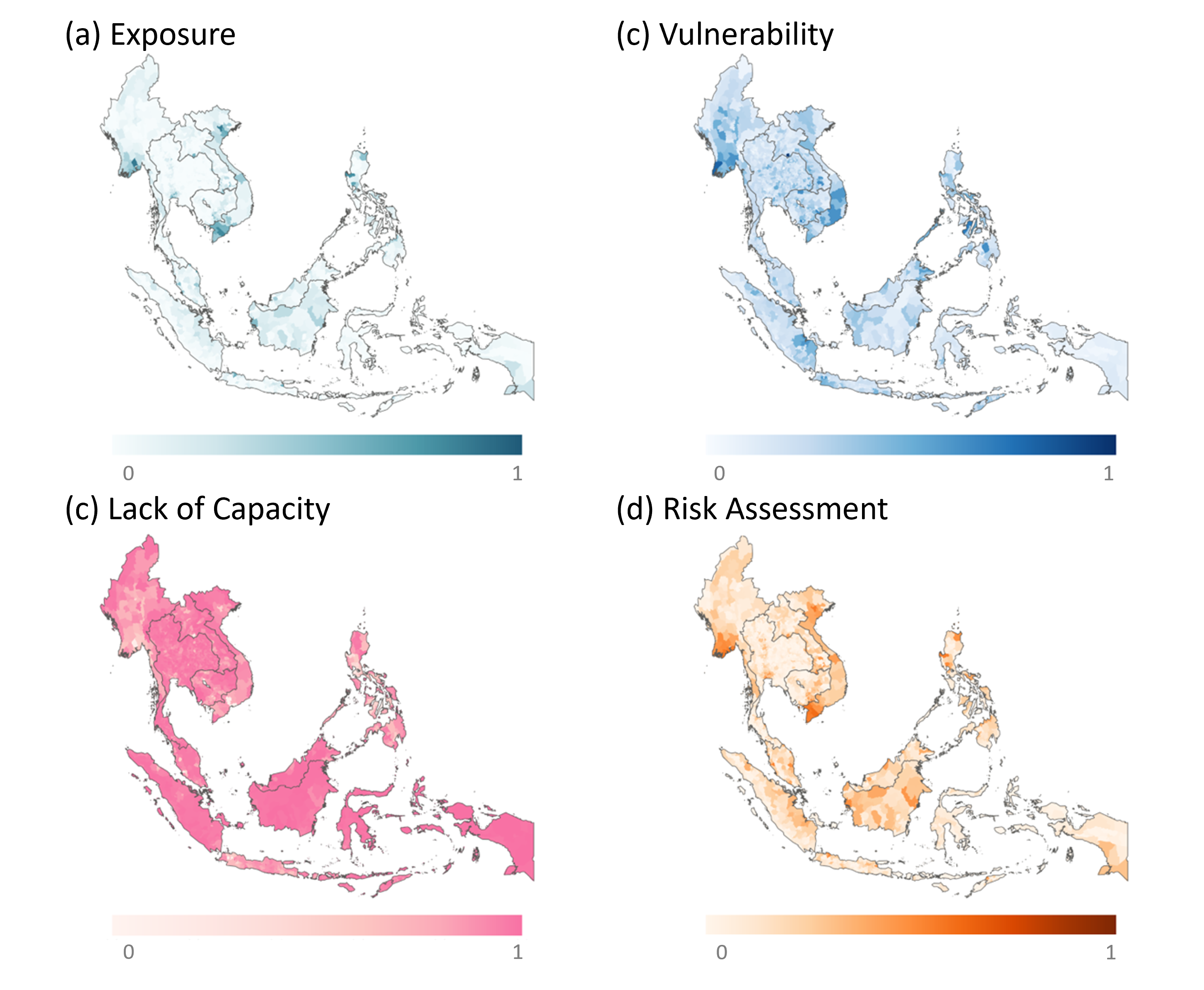}
        \caption{Flood risks in Southeast Asia Countries}
        \label{fig:flooding}
    \end{subfigure}
    
    \caption{Open Disaster Risk Assessment Map for Southeast Asia  reals a regional inequality in term os climate and social justice. Herein, we focus on two common disaster types: Cyclone and Flood. The risk assessment considers demographic, environmental, infrastructure, accessibility, and hazard-related data to support disaster risk and resilience analysis, aggregated at admin level 2.}
    \label{fig:flood_cyclone_SEA}
\end{figure}

Currently, a large amount of GeoAI workflow relies on petabytes of continuous Earth Observation (EO) data. However to process these datasets, especially during time-critical disaster response windows that demand fast and intensive computing, results in massive GPU hours, thus lead to substantial carbon footprint \citep{patterson2021carbon, shi2025geography}. Herein, it is a paradox that the immense energy required to train and deploy these models can actively exacerbates the anthropogenic climate change driving the very disasters, for example intensified hurricanes and prolonged droughts, that GeoAI models are designed to monitor and mitigate. Furthermore, this paradox is compounded by an increasing global inequality and digital divide \cite{tschakert2013inequality, huang2025geospatial}. The impacts of climate change are profoundly uneven, with countries in the Global South are disproportionately vulnerable to extreme weather events. For examples, Figure \ref{fig:flood_cyclone_SEA} demonstrate an inequality of flood and cyclone risk assessment among Southeast Asia (SEA) countries, where open risk assessment datasets are made available n the Humanitarian Data Exchange (HDX) by HeiGIT at Heidelberg University \footnote{https://heigit.org/open-risk-assessment-datasets-for-easier-humanitarian-action/}. Therefore, given the high financial and computational costs associated with operating those energy-intensive GeoAI systems, there is a severe barrier for Global South nations, making it exceedingly difficult for them to afford the advanced disaster risk reduction and mapping technologies they desperately need.

A commitment to sustainability within responsible GeoAI requires moving away from purely performance-based models toward Green AI and carbon-aware computing \citep{schwartz2020green}. This shift is critical not just for the environment, but for climate social justice. Instead of training massive, energy-hungry models from scratch during every disaster, stakeholder can reuse approaches like Retrieval-Augmented Generation (RAG) to pull from existing GeoAI models, saving significant computing power. Moreover, localized edge computing can also play a role by processing data near the disaster site, cutting down on transmission loads and the need for expensive cloud servers \citep{xu2020big, ijaz2023uav}. By adopting these Green AI principles, the GIS community can align its technical progress with net-zero climate goals while ensuring that resource-constrained regions can actually afford to use AI-based diaster response systems \citep{bolon2024review}.

\subsection{Ethics}

GeoAI is being integrated into disaster mapping at a pace that often leaves ethical considerations trailing behind \citep{li2024geoai, oluoch2024crossing, chen2026infographic, zhao2026geoai}. Location data is intensely personal; a single spatial footprint can expose an individual's identity, habits, and socioeconomic status \citep{schlapfer2021universal}. When a climate extreme or disaster strikes, local stakeholders rely on a surge of crowdsourced mapping, mobile phone telemetry, and high-resolution RS imagery to locate affected population and to estimate disaster damage \citep{deng2021high, li2025cross}. But this data collection happens in a time-critical context. As physical infrastructure collapses, institutional ethical safeguards usually fail right alongside it, leaving already marginalized groups completely exposed to higher vulnerability and risk \cite{kwan2005emergency, kwan2012gis}.

Even if the immediate goal of disaster mapping is to deliver life-saving aid, failing to apply privacy protections (e.g., spatial k-anonymity or trajectory aggregation \citep{mckenzie2023privacy}) can easily reveal individual privacy and lead to further ethical issues. Beyond privacy, GeoAI models can risk automating spatial redlining; because disaster models tend to inherit the inequalities already built into our cities, a triage approach trained on historical data will naturally prioritize privileged, well-documented neighborhoods and communities while leaving behind marginalized and less-privileged communities from receiving critical resources when they need them most. To prevent these compounded ethical risks, it requires a firm commitment to data sovereignty and ethical guidelines, aligning with the commonly accepted CARE principles (i.e., Collective Benefit, Authority to Control, Responsibility, Ethics)\citep{carroll2023care}. Moreover, for the GIS community, responsible GeoAI shall embed ethics into disaster mapping practices by actively mandating privacy safeguards, designing equitable algorithm, and providing constant human oversight, to ensure that GeoAI mitigates the human cost of climate extremes rather than adding to it.

\section{Towards A Conceptual Governance Model for Responsible GeoAI}

\begin{figure}[!ht]
    \centering
    \includegraphics[width=0.9\linewidth]{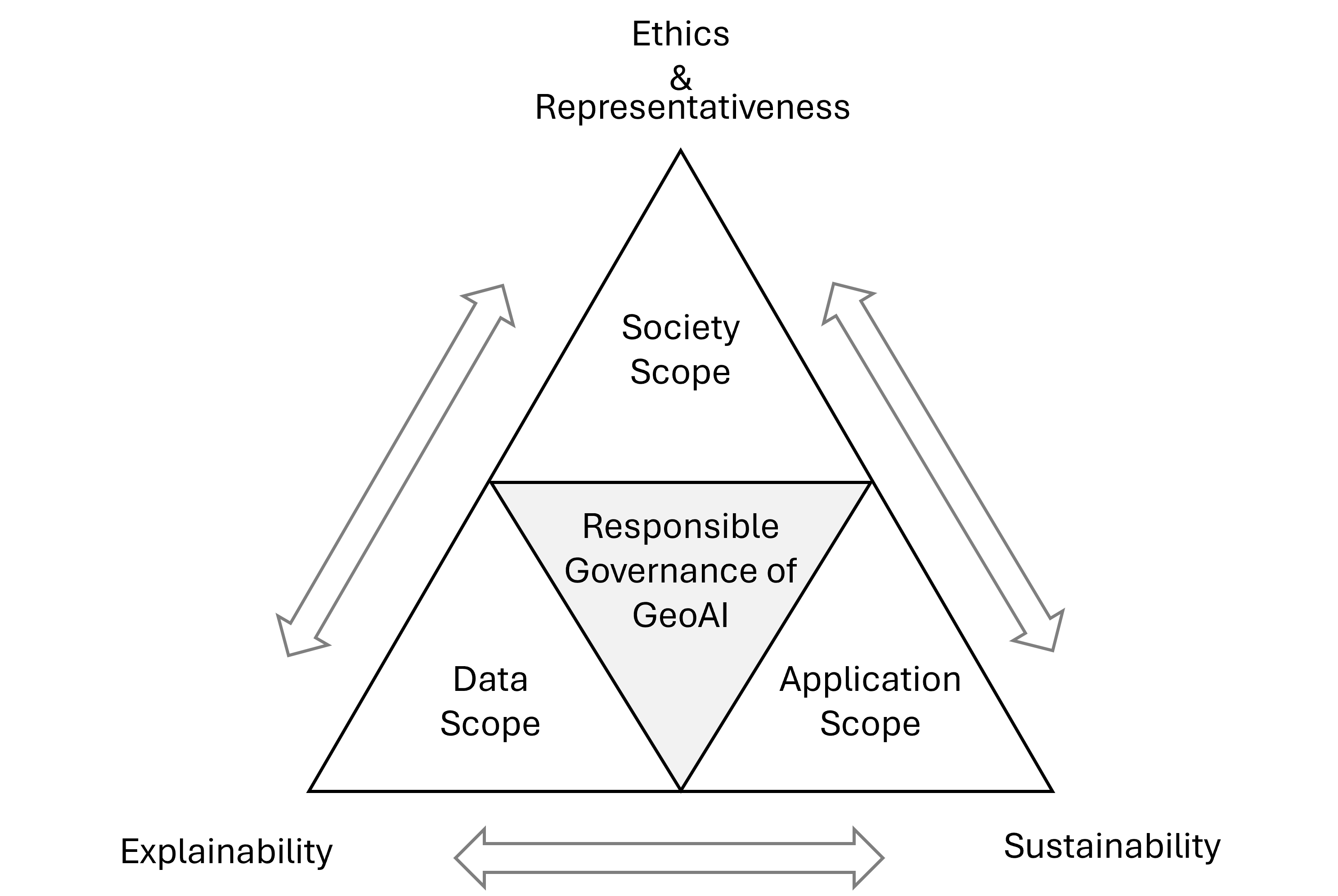}
    \caption{A conceptual governance model for Responsible GeoAI, including three distinct scopes: data, application, and society scopes, bringing together the aforementioned four dimensions (i.e., Representativeness, Explainability, Sustainability, and Ethics).}
    \label{fig:concept_Model}
\end{figure}

Besides the nexus, preliminarily through Representativeness, Explainability, Sustainability, and Ethics, the operationalization of Responsible GeoAI in the chaotic, time-critical scenarios of disaster response however remains challenging. To bridge this gap between theories and real-world practices, we presents a conceptual governance model for Responsible GeoAI (Figure \ref{fig:concept_Model}), which is inspired by \citep{abraham2019data, li2024geoai}, showing how to governance AI-driven geospatial technologies (e.g., across climate extreme and disaster mapping contexts).

To facilitate the responsible governance of GeoAI, this Model defines three operational scopes:

\textbf{Data Scope:} To govern big geospatial data, for instance multi-temporal Earth Observation (EO) imagery for flood extent mapping \citep{wagner2026fully} or crowdsourced mobile telemetry used to track displaced populations \citep{zhou2024understanding}. The data scope in responsible GeoAI mandates advanced protocols for data quality, privacy, and metadata standardization to ensure a responsible workflow. For instance, in a post-hurricane mapping, efficient data triage protocols should be enforced to enhance security of private data and prevent the ingestion of crowdsourced data that systematically skews toward privileged, digitally connected neighborhoods.
    
\textbf{Application Scope:} Herein, to navigate the deployment and algorithmic execution of GeoAI models, such as real-time wildfire monitoring systems \citep{badhan2024deep} or post-disaster building damage classifiers \citep{li2026smart}, it requires standardized and autonomous GIS workflow, including data integration strategies, progress monitoring, and quality control protocols to ensure reliability during active crises. For example, if a flood mapping algorithm begins with misclassifying dense cloud shadows as inundated areas, the governance system must dictate how the GeoAI model is immediately audited, calibrated, or checked by human experts.
    
\textbf{Society Scope:} The third aspect addresses the human, community, and policy dimensions of responsible GeoAI governance \citep{kwan2012gis, nelson2022accelerating}. It necessitates the clear allocation of decision-making authority, such as explicitly defining whether a local emergency management agency, a humanitarian NGO, or a private insurance company may act as the primary data owner during a disaster. Beyond this, ethical Models like transparent communication and interdisciplinary training are essential to foster a responsible GeoAI culture, ensuring that stakeholders are fully equipped to interpret GeoAI-generated risk maps critically without over-relying on the AI systems.

These three operational scopes interact continuously, anchored by the four dimensions of responsible GeoAI. First, the Society \& Data scopes is grounded by \textbf{Ethics \& Representativeness}, ensuring that climate and disaster data harvesting respects vulnerable communities and reflects real-time information without encoding historical infrastructural biases. Second, \textbf{Explainability} bridges the Data and Application scopes by translating multimodality geospatial big data (e.g., VGI, RS, and social media, etc.) into interpretable decision-making insights. This facilitates local decision makers to understand exactly why a specific neighborhood or community is flagged for evacuation and rescue. Finally, \textbf{Sustainability} brings together the Application-Society scopes, ensuring the computational footprint of GeoAI aligns with long-term climate resilience rather than exacerbating the very extremes it monitors. Last but not least, this conceptual Model helps to convert responsible GeoAI from theoretical discussion to accountable, operational GeoAI governance in climate and disaster risk reduction practices.

\section{Conclusion}

As the frequency and intensity of climate extreme events increase, the integration of GeoAI into disaster mapping offers stimulating potential in supporting disaster response at scale. However, as this position paper argues, predictive performance alone is fundamentally insufficient for such time-critical, life-saving practices. The bold and irresponsible deployment of GeoAI risks amplifying inherent spatial inequalities, preventing effective emergency decision-making, and resulting in severe environmental costs that eventually contradict global climate migration goals. 

To theoretically address these challenges, this position paper unbox the nexus of responsible GeoAI through four interrelated dimensions, namely Representativeness, Explainability, Sustainability, and Ethics. Bearing in mind, we then propose a conceptual governance model for responsible GeoAI. By classifying operational governance practices into Data, Application, and Society scopes, our conceptual governance Model offers a high-level theoretical guideline to safeguard responsible deployment, ensure decision reliability, and mandate carbon-aware Green AI, especially in climate extreme and disaster mapping applications. 

In conclusion, the future of climate extreme and disaster risk reduction does not lie solely in creating numerous GeoAI models, but in fostering a responsible culture and ecosystem where AI empowers geographical reasoning and local community context. Looking into the future, this position paper calls for the action from the GIS community to proactively embed responsibility into the very fabric of GeoAI algorithms and operational workflows. In this context, we believe that by shifting from purely performance-driven evaluation to responsible, ethical, and sustainable governance can we ensure that GeoAI genuinely contributes to a disaster resilient future for our human society, particularly for the less privileges but more climate vulnerable Global South.

\section*{Acknowledgments}

We thank Prof. A-Xing Zhu for the valuable comment and suggestion in the early stage of this work. We acknowledge the support in the part of the visualization by Zhihang Liu from CUHK. This work was supported by the Start-Up Grant (SUG) project “Geospatial Artificial Intelligence for Climate Resilient Urban Environment” from the National University of Singapore (E-109-00-0036-01).

\section*{Disclosure statement}

The authors report there are no competing interests to declare.

\bibliographystyle{tfv}
\bibliography{interacttfvsample}

\end{document}